# Break Times: Virtual Reality Art Therapy

Yi Rou Yap[1], Yunli Lee[2]

[1] *School of Engineering and Technology, Sunway University, Selangor, Malaysia*

[2] *Research Centre For Human-Machine Collaboration (HUMAC), School of Engineering and Technology, Sunway University, Selangor, Malaysia*

(Email: angelyap1225@gmail.com)

**Abstract ---** **This paper presents a Virtual Reality (VR) art therapy known as "Break Times" which aims to enhance students' mental well-being and foster creative expression. The proposed "Break Times" application mimics the art therapy sessions in the VR environment design. Pilot user acceptance test with 10 participants showed a notable reduction in stress levels, with 50% reporting normal stress levels post-intervention, compared to 20% pre-intervention. Participants praised the "Break Times" therapy's functionality and engagement features and suggested improvements such as saving creations, incorporating 3D painting, and expanding the artmaking scene variety. The study highlights that VR art therapy has potential as an effective tool for stress management, emphasizing the need for continued refinement to maximize its therapeutic benefits.**

## 1 INTRODUCTION

Mental health issues have become increasingly prevalent among various demographics in recent years, particularly adolescents and young adults who face emotional regulation challenges, academic pressure, and societal pressure [1]. Anxiety, depression, and stress are common afflictions that necessitate effective and accessible therapeutic interventions to promote mental well-being and prevent the escalation of these conditions [2, 3]. Research works of [4, 5] underscore the potential of art therapy to address these mental health disorders.

Art therapy has emerged as a promising intervention for addressing various mental health disorders. This therapeutic modality encourages individuals to engage in creative activities such as painting, music, dance, and writing [6]. Art therapy's efficacy lies in its ability to address psychological, behavioral, and physiological disorders by offering a holistic approach to emotional expression and self-reflection, thereby alleviating depression and anxiety [1, 7]. The therapeutic effects of music, as highlighted in [8], emphasize its capacity to infuse energy, stimulate self-healing potential, and foster individuals' confidence.

With the advancement of Virtual Reality (VR) technology, various fields, including healthcare, have been revolutionized by providing immersive and interactive experiences. The integration of VR into therapeutic practices presents an innovative approach to enhancing mental health interventions [9, 10]. VR's immersive nature transcends physical boundaries, offering individuals a transformative space for creative expression and therapeutic engagement. By creating an immersive virtual environment where users can interact with digital art tools, VR art therapy can enhance the therapeutic process, making it more engaging and accessible to adolescents [11, 12].

Since VR has emerged as a promising technology in therapeutic settings, particularly in art therapy. This has motivated us to design "Break Times" - a therapeutic VR environment focused on enhancing mental well-being through art therapy. "Break Times" is designed as a soothing immersive virtual space that allows users to engage with digital art tools, fostering creative engagement. Emphasizing therapeutic elements, it features tailored tools for relaxation and self-reflection using 2D sandbox concept with colors and musical notes. The visual art and melodies composed during the art-making process in "Break Times" are intended to enhance therapeutic experience through creative and sensory engagement. This application serves as a valuable tool for college students, particularly adolescents and young adults, who is seeking a unique and immersive way to take a break and express themselves.

Additionally, a modified set of evaluation questions is designed to assess the impact of the "Break Times"

application on students' mental health. Pilot study was conducted to measure and analyze the stress level of the participants. By merging innovative technology with art therapy principles, the proposed "Break Times" serves as a user-friendly platform and provides a transformative experience by contributing to the convergence of virtual reality and mental health well-being.

## 2 RELATED WORK

L. Shamri Zeevi [12] documented case studies where adolescents with social and body image issues benefited from VR art therapy. The results shown: Eric, a 16-year-old with social and academic difficulties, and Alma, a 13-year-old with body image issues, both found VR a valuable medium for self-expression and therapeutic engagement. Besides that, G. Kaimal et al. [10] explored VR as a visual guide for relaxation techniques, while I. Hacmun et al. [13] examined VR's impact on the therapist-client relationship, noting both challenges and benefits. Studies highlight its potential in fostering a secure, expressive environment, offering individuals control and facilitating self-expression [14]. VR also presenting benefits like stress reduction and cognitive enhancements in neurorehabilitation [4, 2, 15, 16].

We have conducted research on break effectiveness during mentally demanding tasks has mixed results towards art therapy. Longer breaks may lead to an unsustainable initial burst of performance, followed by a decline. Conversely, micro-breaks consistently enhance well-being, increasing vigor and reducing fatigue [18, 19]. Besides, [17] presented that the art therapy sessions are unique, tailored to individual client needs and therapeutic goals. Sessions typically include four components: "Check-in", "Artmaking", "Discussion/Verbal Processing", and "Closure". The "Check-in" phase updates the client's status and goals. During "Artmaking," clients engage in unstructured or structured activities, fostering self-direction or following guidance. The "Discussion/Verbal Processing" phase uses artwork as a metaphor to explore emotions and situations, particularly in trauma cases. The final phase, "Closure," helps transition back to daily life with grounding exercises.

G. J. Riew et al. [6] emphasizes the importance of music in VR art therapy, suggesting research into patient preferences and optimal integration methods to enrich multisensory interventions. In addition, Q. X. Sun et al. [20] discussed the therapeutic benefits of sandbox games like "Minecraft," which offer a safe space for self-expression, relaxation, and psychological growth,
addressing issues such as low self-esteem and emotional instability [21]. Furthermore, [22] explores the impact of color therapy on emotions and academic stress, demonstrating the calming effects of colors like blue and yellow to high school students. These advancements are transforming VR art therapy, offering diverse and effective multisensory interventions to enhance emotional and psychological well-being.

## 3 "BREAK TIMES" - VR ART THERAPY DESIGN

We mimic the four art therapy session components to design "Break Times" VR art therapy application. First, after launched art therapy with "Start" button, the "Scenario Selection" as "Check-in" component will be displayed. This component allows users to customize the therapy session based on their needs through three options. The "Level Selection" option enables users to choose from quick, moderate, or long breaks. Based on the literature findings, we proposed three reasonable break schedules as level options with alert timer (quick: 5 mins, moderate: 15 mins, long: 25 mins) to sustain artmaking productivity and well-being. The "Randomized" option introduces an element of surprise by selecting a drawing scenario at random mode. The "All Scenarios" displays all available drawings and offers a comprehensive view for users to select their preference therapy session.

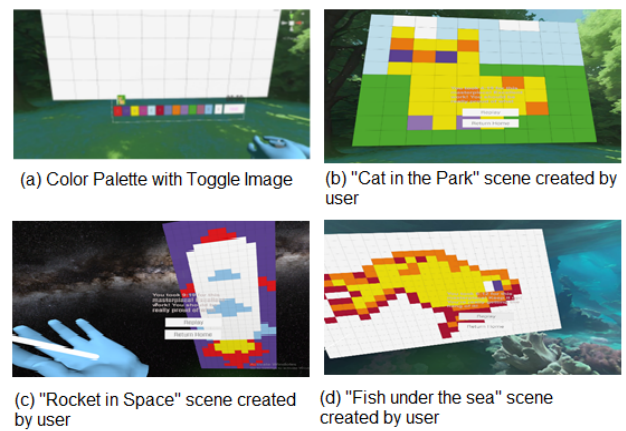

Fig.1 "Break Times" the VR Therapy Interface and Scenes

The second component is "2D Therapy Scene" as "Artmaking" session. The 2D sandbox therapy elements were incorporated to allow users to freely create and manipulate their virtual environment, fostering self-expression and creativity. In addition, chromesthesia (chromotherapy) was integrated by linking musical notes to corresponding colors, it providing a synergistic balance between music and art. Each 2D Therapy Scene adjusts the drawing board size and scenario based on break length, with options like "Cat in the Park" for quick breaks, "Rocket in Space" for moderate breaks, and "Fish Under the Sea" for long breaks. Figure 1 (a) shows the "Cat in the Park" scene with the color palette that linked to the corresponding musical note. The palette moves with the user's view, and an eraser button allows for corrections. A toggle image feature is available for users who seek guidance or inspiration for their creative drawings. This multisensory approach aims to stimulate both visual and auditory senses, enhancing emotional and psychological well-being during art therapy sessions.

The third component is "Completion" as "Discussion/Verbal Processing". Upon completion, the system calculates a score based on the time taken and the number of blocks colored, followed by musical tunes with encouraging messages tailored to their performance as shown in Figure 1 (b) to (d) on respective scene. Users can choose to return to the Main Menu or replay their artwork. The replay feature re-enacts coloring actions in sequence, coloring the blocks every 0.4 seconds while playing the corresponding musical tunes. The last component is "Simple Chat Box" as "Closure", this will be returning to the Main Menu, where users are prompted to share their emotional state, promoting engagement with their emotional well-being.

## 4 EVALUATION AND RESULTS

### 4.1 Preliminary Study

A survey was conducted to assess the impact of VR artmaking on stress reduction among young adults (ages 18-25) in Malaysia using Google form. A total of 68 responses were collected primarily from various national and private universities in Malaysia. Most participants held bachelor's degrees, with a significant number in IT, Business, and Engineering fields. Notably, 35.3% had no prior VR experience, and 54% had not recently engaged in artmaking. Participants preferred VR art mediums such as 2D painting and 3D modeling. These results highlight a need for an accessible introduction to VR art therapy, catering to users with limited prior experience.

### 4.2 User Acceptance Testing

Ten (10) university students aged 18 and above (7 males and 3 females) were selected to test the "Break Times" VR art therapy game under controlled conditions at the lab. The participants stress levels were measured using a modified DASS-21 survey that focused on stress-related questions (only 7 questions) before and after the VR session. Scores were compiled and evaluated to assess relative stress levels. The post-test survey was extended with three additional sections: functionality and technical performances, user experience and engagement, and an open-ended question on the proposed solution.

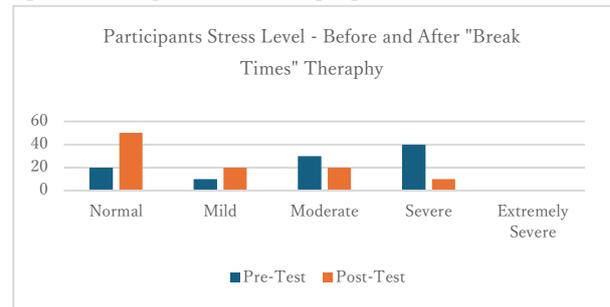

Fig.2 Pre-Test vs Post-Test: Participants Stress Level

Pre-Test showed 80% of participants exhibited abnormal stress levels, highlighting a significant stress prevalence within participants. This underscores the need to address stress-related issues effectively, given its potential impact on mental and physical health, productivity, and quality of life. Upon completion of "Break Times" VR art therapy, 50% of participants still had abnormal stress levels, but there was a notable 30% reduction in severe stress levels as shown in Figure 2. The reduction indicates that the "Break Times" VR art therapy effectively helps manage and alleviate high stress.

In the extended post survey, participants generally rated the application's functionality and technical performances positively, with high scores for user experience and engagement as shown in Figure 3. The functionality and technical performance with most rating it at 4 or 5. Though minor technical issues such as occasional bugs and slight lag were noted by a few users, the loading time was acceptable for the majority, with some suggestion improvements for a more seamless experience.

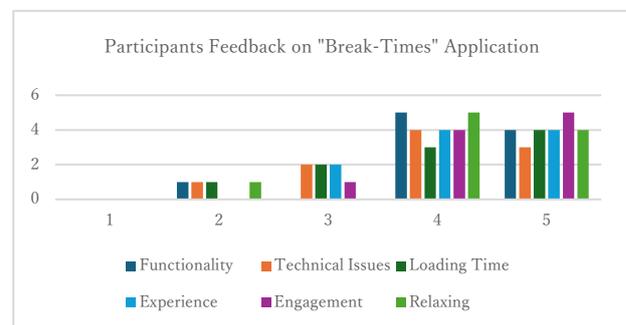

Fig.3 Participants Feedback on the Functionality, Technical Performances, Experience, Engagement and Relaxing

Although most rating for engagement and relaxation levels is either 4 or 5, however, there is a small number of participants rated their relaxation at level 2 indicating that a more personalized approach may be beneficial for stress relief. Overall, the feedback highlights the therapeutic benefits of the application while pointing out areas for future technical optimization and personalized user experiences such as save creations, the addition of 3D painting, and new game types.

## 5 Conclusion

The "Break Times" VR art therapy game has proven to be an effective tool for stress reduction and creative therapy. Results indicate that the application successfully reduces stress and fosters creativity, with positive feedback across various break lengths. The immersive environment, complemented by soothing audio and interactive elements, enhances its therapeutic benefits and versatility. Nevertheless, several areas for improvement have been identified such as the save feature, adding 3D painting capabilities, and adding a variety of artmaking scenes with therapeutic activities will help prevent user monotony and maintain engagement.